\pdfoutput=1
\documentclass[aps,prl,twocolumn,letterpaper,showpacs,preprintnumbers,amsmath,amssymb]{revtex4}
\usepackage{graphicx} 
\usepackage{bm}       
\usepackage{amsmath}
\usepackage{SIunits}  
\usepackage{color}

\begin{document}

\title{Cluster assimilation and collisional filtering on metal-oxide surfaces}

\author{Daniel~A.~Freedman}
\author{T.A.~Arias}
\affiliation{Laboratory of Atomic and Solid State Physics, Cornell University, Ithaca, NY 14853}

\date{\today}

\begin{abstract}
We present the first {\em ab initio} molecular dynamics study of
collisions between metal-oxide clusters and surfaces. The resulting
trajectories reveal that internal degrees of freedom of the cluster
play a defining role in collision outcome. The phase space of incoming
internal temperature and translational energy exhibits regions where
the collision process itself ensures that clusters which do not
rebound from the surface assimilate seamlessly onto it upon impact.
This filtering may explain some aspects of recent observations of a
``fast smoothing mechanism'' during pulsed laser deposition.
\end{abstract}



\maketitle

The success of pulsed laser deposition of complex oxide films raises
the fundamental question of how a disordered distribution of incoming
clusters incorporates seamlessly into highly ordered crystals with
complex unit cells.  In pulsed laser deposition, a laser pulse
impinges upon a target and ejects hot material into a plasma plume
which then condenses upon a growing substrate\cite{willmott}.  With
empirical tuning of parameters such as laser wavelength and energy
density, laser pulse width and separation, partial pressure of the
ambient gas, and substrate temperature, the resulting film can be made
to grow smoothly and nearly defect free\cite{ChriseyHubler}.  Little
is known about what underlying fundamental processes these external
parameters control.  In recent years, sub-second time-resolved {\em in
situ} x-ray measurements of growth by this process have become
possible\cite{Eres2002,FleetBrock}.  These experiments suggest that
incoming clusters from the plume are incorporated into the substrate
on sub-millisecond time scales in a ``fast smoothing mechanism'' which
occurs too quickly to be explained by traditional diffusional
smoothing\cite{FleetBrock,Willmott2006}.  The present work addresses
the question of whether smoothing mechanisms exist over the time scale
of the actual {\em collisions} between the clusters and the surface.

To address the existence of such collisional mechanisms, atomistic, as
opposed to continuum, descriptions of growth are most appropriate.
Current atomistic studies of crystal growth employ a number of
methods, such as kinetic Monte
Carlo\cite{Kotomin,Lam2002,PomeroyBrock}, molecular dynamics based on
classical interatomic potentials\cite{Zhou2006,Aguado,KuboJCP},
combinations of the two\cite{JacobsenSethna,PomeroySethna},
accelerated molecular dynamics\cite{Voter}, and {\em ab initio}
calculations of already deposited
material\cite{galli,musolino1999,Barcaro2005}.  However, kinetic Monte
Carlo methods, by their nature, handle only diffusive events, not
actual deposition dynamics.  Although classical molecular dynamics can
address collisions, the use of interatomic potentials raises the issue
of accuracy, particularly in oxides, which generally have a number of
different atomic species and complex physical chemistry.  While {\em
ab initio} studies of already deposited material give insights into
metastable structures and transition states, they do not do so for the
kinetic mechanisms active during collisions.  To study such
mechanisms, this {\em Letter} presents the first {\em ab initio}
molecular dynamics calculations of collisions between metal-oxide
clusters and surfaces, with magnesium oxide selected as a simple model
system.

{\em Methods ---} All density-functional theory\cite{hohenbergkohn}
calculations below employ the local density
approximation\cite{kohnsham} and use the total-energy plane-wave
pseudopotential technique\cite{payne} with a 20~hartree (20~H) cutoff.
The pseudopotentials include non-local corrections of the
Kleinman-Bylander form\cite{kleinmanbylander} for the {\em p} and {\em
d} channels.

We represent the MgO (001) surface with a $3\times3$ periodic
supercell and three layers of atoms.  The surface slabs are separated
by 12.10~\AA\ or 18.15~\AA\ of vacuum for cold and hot incoming
clusters, respectively.  The in-plane lattice constant of the
supercell (8.55~\AA) corresponds to that of the relaxed bulk crystal.
Finally, we integrate over the Brillouin zone for this wide band-gap
insulator using a single k-point at the zone center.

For molecular dynamics, we employ the Verlet algorithm\cite{verlet}
using a time step of 2.04~fs.  We maintain the electrons within 0.1~mH
of the Born-Oppenheimer surface using preconditioned
conjugate-gradients within the analytically continued functional
approach\cite{ariasjoan}.  These parameters conserve total energy to
within 3~mH (0.3\% of the collision energy) throughout.  The initial
condition of the slab is taken to be its fully-relaxed vacuum
configuration at zero temperature, with the bottom layer fixed at bulk
locations, a constraint also maintained during the molecular dynamics.

To study incorporation of clusters beyond the size of simple molecular
units, we study collisions with the eight-atom ($N_\text{cl}\equiv 8$)
cubic ``magic cluster''\cite{delaPuente1997}.  To explore the role of
internal degrees of freedom, we consider otherwise identical
collisions with both ``cold'' and ``hot'' incoming clusters.  We
prepare the cold cluster by full relaxation in vacuum and the hot
cluster by adiabatic heating (with no net momentum or angular
momentum) to an internal kinetic energy of $K_\text{int}=0.057$~H.
This energy corresponds to an internal temperature of
$T=K_\text{int}/(3 N_\text{cl} k_B/2)\sim 1500$~K, well below the bulk
melting temperature, both {\em ab initio} (3110~K\cite{alfe}) and
experimental (3250~K\cite{ronchisheindlin}).  Finally, we give the
incoming clusters a translational kinetic energy representative of the
range ($\sim$10 to $\sim$100~eV) which yields smooth growth in
energetic deposition\cite{lowndes, willmott}.  In particular, we
choose 1~H ($\approx$ 27.21~eV), near the geometric mean of this
range.

\begin{figure}
  \begin{centering}
    \includegraphics[width=3.3in]{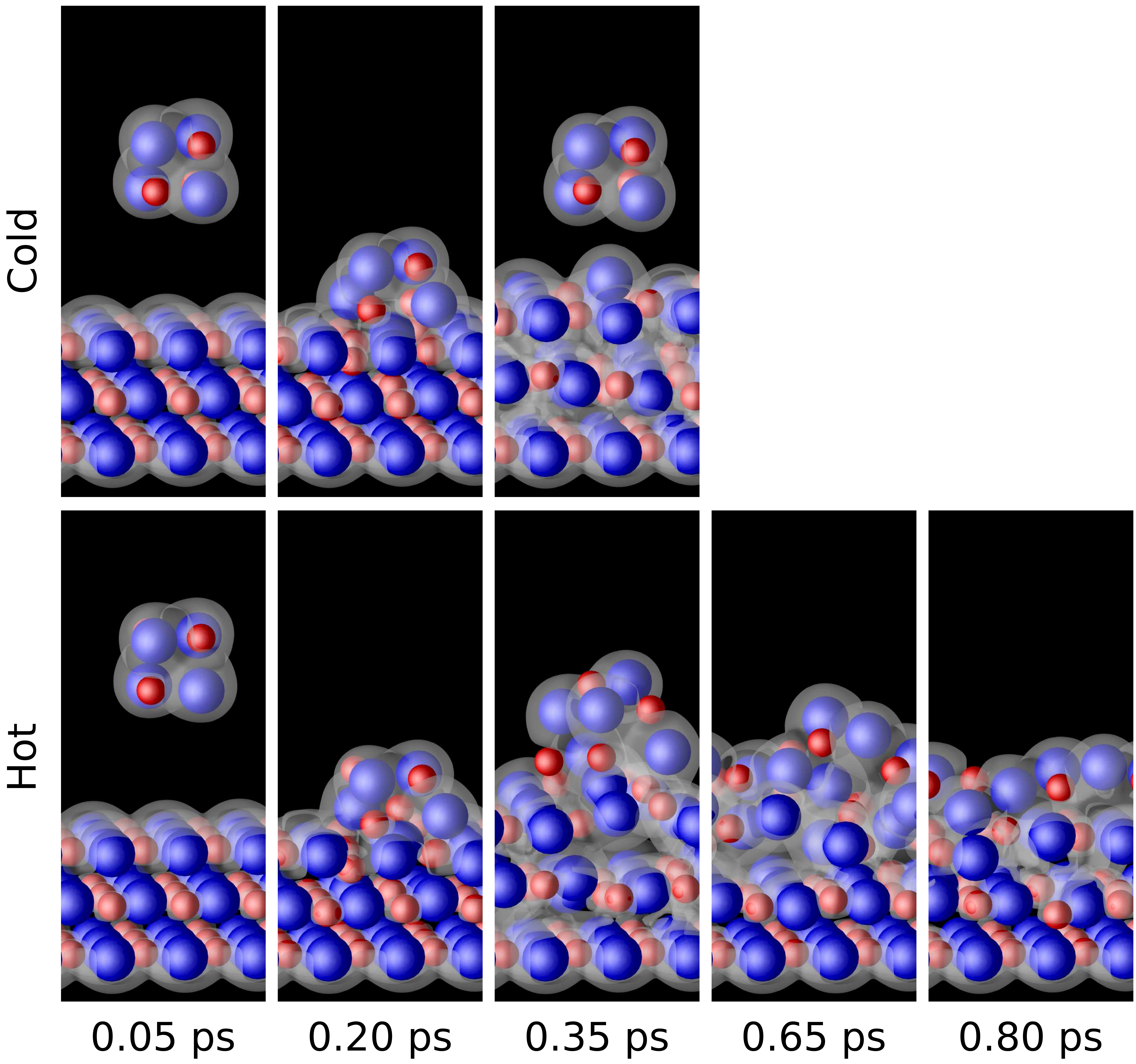}
    \caption{(color) {\em Ab initio} molecular dynamics snapshots of
      hot and cold clusters in collision with magnesium-oxide surface:
      Mg cores (red), O cores (blue), electron-density isosurfaces
      (white), coordinate axes (upper right panel).}
    \label{fig:Snapshots}
  \end{centering}
\end{figure}

{\em Results ---} Figure~\ref{fig:Snapshots} presents snapshots of our
raw results at representative times.  The central result for the cold
cluster (top row) is that it rebounds and does not bind to the
surface.  Initially ($t=0.05$~ps), the cold cluster approaches the
surface.  After contact of the electron clouds, the cluster compresses
while pushing atoms on the surface into the slab ($t=0.20$~ps).  The
cluster then rebounds into the environment ($t=0.35$~ps) and does not
contribute to growth of the surface.

In marked contrast, the hot cluster binds to the surface.  Initially
($t=0.05$~ps), it approaches the surface with $\sim$1500~K of internal
kinetic energy and the same impact parameters and velocity as the cold
cluster.  After contact, this cluster also compresses while pushing
atoms on the surface into the slab ($t=0.20$~ps) in a configuration
quite similar to that of the cold cluster.  This collision dissipates
sufficient translational kinetic energy for the cluster to bind to the
surface ($t=0.35$~ps).  Bound, the cluster equilibrates with the
surface until it assumes a rock-salt configuration conforming to the
underlying crystal ($t=0.65$~ps).  Thereafter, the cluster cools until
thermal vibration accounts for the remaining distortions
($t=0.80$~ps), at which point the cluster has assimilated seamlessly
onto the underlying surface.

\begin{figure}
  \begin{centering}
    \includegraphics[width=3.5in]{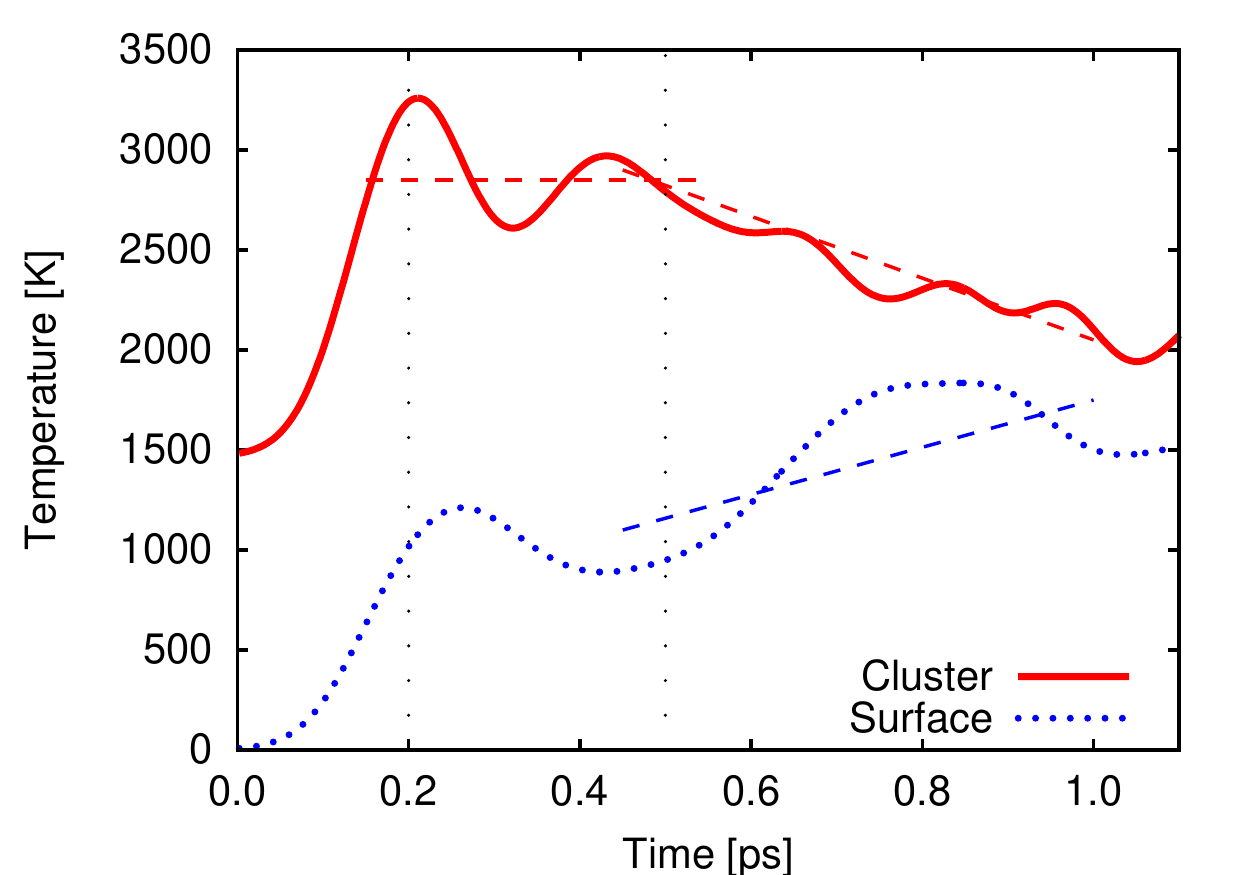}
    \caption{(color online) Cluster (solid curve) and surface (dotted
      curve) temperature versus time: impacting, freezing, cooling
      intervals (left to right, demarked by dashed vertical lines).
      Dashed horizontal and diagonal lines are guides to the eye.}
    \label{fig:HotTempCurve}
  \end{centering}
\end{figure}

To aid interpretation of the snapshots of the hot cluster,
Figure~\ref{fig:HotTempCurve} presents the internal kinetic energy
(expressed as a temperature) of the hot cluster and the slab as a
function of time.  To reduce the appearance of fluctuations associated
with the small numbers of atoms, we convolve the data with a Gaussian
of width 0.05~ps.  Impact occurs in the first interval indicated in
the figure ($t<0.20$~ps), during which the temperature of the slab and
cluster both rise.  In the following two intervals
(0.20~ps$\;<t<\;$0.50~ps; 0.50~ps$\;<t$), the temperature of the slab
rises consistently.  In contrast, the temperature of the cluster
oscillates during the second interval (0.20~ps$\;<t<\;$0.50~ps) and
then drops consistently toward that of the slab during the final
interval (0.50~ps$\;<t$).  Finally, we note that a similar analysis
shows that the temperature of the cold cluster also rises during the
collision, but only to approximately 2000~K, well below the bulk
melting point of $\sim3100$~K.

{\em Discussion ---} During its collision, the cold cluster never
reaches the melting point and, recovering its initial form, is able to
convert enough of the energy stored during the impact back into
translational motion to escape the surface.  In contrast, the hot
cluster approaches the melting point on impact and loses its original
structure.  Thereafter, it cannot convert sufficient energy from the
impact into translational motion to escape the surface.

Despite the small number of atoms in the cluster, thermodynamic
concepts provide a useful framework for these observations.  If we
describe the disordering of the hot cluster as ``melting,'' then the
translational energy which the hot cluster is unable to recover is
analogous to the latent heat of fusion.  We then expect the melted
cluster to equilibrate with the underlying surface --- first by
maintaining a constant temperature while releasing the heat of fusion
into the surface, and then by cooling until its temperature matches
that of the surface.

The internal temperature of the hot cluster
(Figure~\ref{fig:HotTempCurve}) arguably shows just this behavior.
During the impact in the first interval ($t<0.20$~ps), the internal
energy of the hot cluster rises to approach the bulk melting point
(3100~K).  In the second, ``freezing,'' interval
(0.20~ps$\;<t<\;$0.50~ps), the temperature shows fluctuations around a
relatively constant value (horizontal dashed line in
Figure~\ref{fig:HotTempCurve}).  Consistent with this, the snapshot
from the center of this interval ($t=0.35$~ps) shows a cluster in the
midst of changing its topology to conform to the underlying surface.
In the third interval (0.50~ps$\;<t$), the cluster cools while the
surface heats as the two equilibrate to a common temperature.
Snapshots from this interval ($t=0.65$~ps and $t=0.80$~ps) show the
cluster assimilated into the surface, with only thermal vibrational
motion remaining.

Within the preceding framework, we make the following predictions.
First, the temperature of a cluster which ensures melting upon impact
should decrease with increasing incoming translational energy.  There
thus should be a {\em melting curve} in the phase space of incoming
cluster temperature versus incoming translational energy, above which
the cluster melts upon impact.  This curve has a horizontal asymptote
at the melting temperature of the cluster for low translational
energies, with a horizontal intercept where the translational energy
alone is sufficient to melt the cluster.  Second, the cluster
temperature which ensures binding after impact should increase with
increasing translational energy because internal temperature promotes
absorption of mechanical energy.  There thus should be a {\em binding
curve} in the phase space above which the cluster binds.  This curve
has a horizontal intercept at the incoming energy below which the
attraction between cluster and surface always suffices to bind the
cluster, followed by a vertical asymptote at the translational energy
above which the cluster never absorbs sufficient energy to allow
binding.  Due to the ratio of bulk to surface bonds in the cluster,
the translational energy above which the cluster always melts should
be greater than the energy below which it always binds, so that, in
general, the melting and binding curves cross.  Finally, we expect
that collisions will begin to disrupt the topology of the surface for
translational energies somewhere near the point where there is enough
energy to melt cold clusters upon impact, thus defining a {\em
disruption curve}.

\begin{figure}
  \begin{centering}
    \includegraphics[width=3.5in]{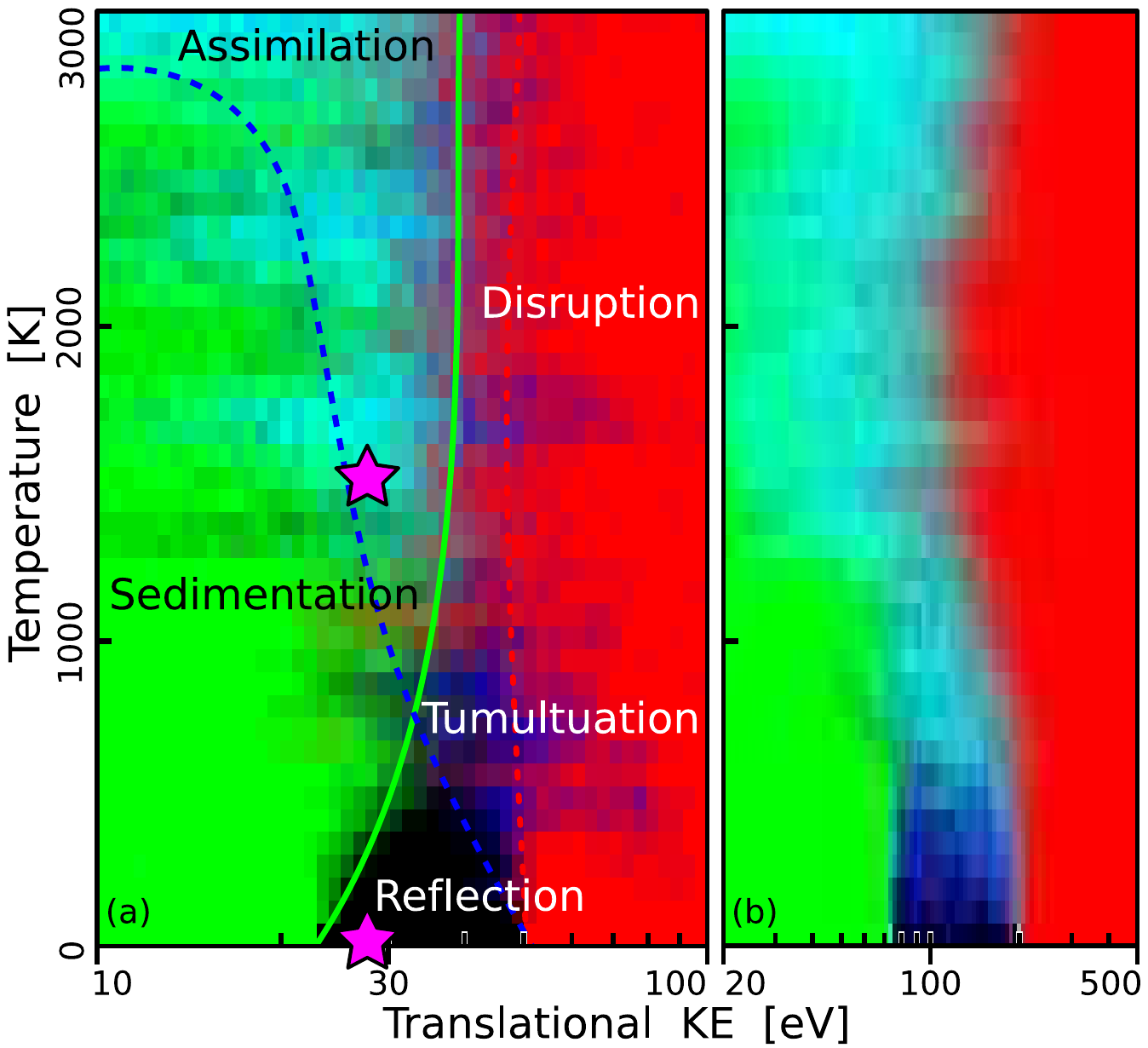}
    \caption{(color) Phase space of incoming temperature versus
      translational energy for (a) small and (b) large supercells:
      {\em ab initio} results (purple stars); proposed {\em melting
      curve} (dashed blue), {\em binding curve} (solid green) and {\em
      disruption curve} (dotted red) overlaid on regions of {\em
      assimilation} (cyan), {\em reflection} (black), {\em
      tumultuation} (blue), {\em sedimentation} (green), and {\em
      disruption} (red), from classical molecular dynamics
      simulation.}
    \label{fig:Phasespace}
  \end{centering}
\end{figure}

Figure~\ref{fig:Phasespace} shows this phase space.  In particular,
Figure~\ref{fig:Phasespace}(a) illustrates a {\em melting curve}, {\em
binding curve} and {\em disruption curve}, and indicates points
corresponding to the above {\em ab initio} molecular dynamics
calculations.  (The specific placement of the curves and background
coloring derive from classical molecular dynamics simulations
described below.) The melting and binding curves define four regions
in the phase space: {\em assimilation}, {\em reflection}, {\em
tumultuation}, and {\em sedimentation}.  In the region above both
curves, clusters bind and melt and subsequently deform and {\em
assimilate} seamlessly onto the surface upon impact.  In the region
below both curves, clusters neither bind nor melt and thus {\em
reflect} intact from the surface.  In the region above the melting but
below the binding curve, clusters melt but do not bind and thus {\em
tumult} from the surface in a deformed state.  In the region above the
binding but below the melting curve, clusters bind but do not melt and
thus {\em sediment} intact on the surface.  Finally, the {\em
disruption} curve defines the region where the surface topology is
disrupted.  The {\em ab initio} results conform to this demarcation of
phase space, with the translational energy of 1~H such that the cold
cluster (lower purple star in Figure~\ref{fig:Phasespace}(a)) reflects
and the hot cluster (upper purple star in
Figure~\ref{fig:Phasespace}(a)) assimilates.

To test whether these conclusions are general and insensitive to
details of the underlying interactions, we construct a simple
interatomic potential model of a generic divalent ionic crystal in the
form of a Coulomb interaction and a pairwise short-range repulsion,
\begin{equation*}
  U = \frac{1}{2}\, \sum_{i\ne j} \left[ \frac{q_i q_j}{r_{ij}}
  \left(1 - \mbox{erfc\,}\frac{r_{ij}}{a_{ij}} \right) + A_0\, e^{-(r_{ij}/a_{ij})^2} \right],
\end{equation*}
where we work in atomic units, $U$ is the total energy of the crystal,
$q_i=\pm 2$ are the ionic charges, $r_{ij}$ is the distance between
atoms $i$ and $j$, $A_0=7$~H (fit to compromise between the lattice
constant and bulk modulus of MgO) measures the strength of the
repulsion between ionic cores, and $a_{ij}$ is a range parameter
defined as the mean ionic radius of atoms $i$ and $j$
($R_{\text{Mg}}=0.66$~\AA, $R_{\text{O}}=1.32$~\AA).  The ``erfc''
term is part of the short-range repulsion between ionic cores and
serves to remove the Coulomb singularity when the ionic cores overlap.
This simple model material prefers rock-salt over the cesium-chloride
structure, as does magnesium oxide, and has a lattice constant and
bulk modulus of 5.0~\AA\ and 240~GPa, both significantly larger (20\%
and 50\%, respectively) than the corresponding experimental quantities
for magnesium oxide.

Using the above model, we repeat the procedure of the {\em ab initio}
molecular dynamics calculations in the same three-layer 3$\times$3
supercell, but now map the phase space in detail: on a grid of 40
values of incoming temperature and 50 values of translational energy,
sampling 25 collisions at each phase-space point, for a total of
50,000 trajectories.  To explore convergence with system size, we also
study a five-layer $6\times6$ supercell using 100,000 trajectories.
Figures~\ref{fig:Phasespace}(a,b) summarize these results.  The blue
intensity of each pixel encodes the probability of melting; the green
encodes that of binding; and the red channel encodes disruption,
overriding blue and green regardless of cluster behavior.  Points
where clusters disrupt the surface thus appear red, whereas points
where clusters assimilate appear cyan, reflect appear black, tumult
appear blue, and sediment appear green. The pixelization of the figure
reflects the discreteness of the sampling and the fluctuations reflect
Poisson statistics ($\pm$20\%).

Remarkably, the data in both Figures~\ref{fig:Phasespace}(a,b)
correspond precisely to the expectations of the physical picture
developed above.  In particular, we find the expected five regions of
behavior separated by the anticipated curves.  Even {\em
quantitatively}, within the corresponding supercell, the {\em ab
initio} results fall correctly into the assimilation and reflection
regions, despite the relatively small area of these regions and the
quantitative differences between the {\em ab initio} and model
materials.  The larger supercell results show the same overall
behavior, with quantitative correspondence for the internal
temperatures and some rescaling of the translational energy.  The
latter effect, we believe, results from a deeper surface providing a
more ``cushioned'' impact.

{\em Conclusion ---} We present the first direct Born-Oppenheimer {\em
ab initio} molecular dynamics calculations to demonstrate that
metal-oxide clusters can assimilate seamlessly onto metal-oxide
surfaces during the collisional time scale ($\sim$1~ps) --- far
shorter than diffusional time scales.  These calculations, along with
extensive classical molecular dynamics simulations and general
physical considerations, support the novel picture that the internal
degrees of freedom of the incoming clusters play an important role in
deposition. The phase space of incoming temperature and translational
energy summarizes important features of collision outcome,
distinguishing regions of melting and binding in terms of curves whose
behavior is easily understood.

The resulting phase diagram leads to new insights into pulsed laser
deposition.  Each laser pulse produces an ensemble of clusters
scattered across the phase space.  For translational energies typical
of experimental conditions for smooth growth, the arrangement of
regions in the phase diagram indicates that the collision process
itself provides an effective filter to ensure assimilation of clusters
onto the surface {\em upon impact}.  For translational energies above
the crossing of the melting and binding curves, incoming clusters
which manage to bind also melt, and thus assimilate rather than
sediment.  We believe that this result may relate to observations of a
``fast smoothing mechanism'' in growth by pulsed laser
deposition\cite{FleetBrock,Willmott2006}.

\begin{acknowledgments}
The authors acknowledge fruitful discussions with J.~Brock, A.~Fleet,
and D.~Dale and primary support from the NSF MRSEC program via grant
DMR-9632275.
\end{acknowledgments}

\end{document}